\documentclass[aip,jcp,amsmath,amssymb,reprint, citeautoscript, floatfix, longbibliography]{revtex4-1}

\usepackage{graphicx}
\usepackage{dcolumn}
\usepackage{bm}

\usepackage[caption=false]{subfig}
\usepackage{textcomp}
\usepackage{booktabs}
\usepackage{hhline}
\usepackage{braket}
\usepackage[dvipsnames]{xcolor}

\DeclareMathSymbol{\mdot}{\mathord}{symbols}{"01}

\newcommand{\erm}{{\rm e}}

\newcommand{\wn}{\,{\rm cm}^{-1}}
\newcommand{\dwn}{\,{\rm D^2}/{\rm cm}^{-1}}

\DeclareUnicodeCharacter{0301}{\'{e}}

\begin{document}

\title[]{Symmetry breaking fluctuations split the porphyrin $Q$ bands}

\author{Zachary R. Wiethorn}
\affiliation{Department of Chemistry, University of Colorado Boulder, Boulder, Colorado 80309, USA}

\author{Kye E. Hunter}
 \affiliation{Department of Chemistry, Oregon State University, Corvallis, Oregon 97331, USA}
  \affiliation{Present address: CRG (Barcelona Collaboratorium for Modelling and Predictive Biology),  Dr. Aiguader 88, Barcelona 08003, Spain}

\author{Andr\'{e}s Montoya-Castillo}
\email{Andres.MontoyaCastillo@colorado.edu}
\affiliation{Department of Chemistry, University of Colorado Boulder, Boulder, Colorado 80309, USA}

\author{Tim J. Zuehlsdorff}
\email{tim.zuehlsdorff@oregonstate.edu}
 \affiliation{Department of Chemistry, Oregon State University, Corvallis, Oregon 97331, USA}

\begin{abstract}
Porphyrins offer a malleable and cost-efficient platform to sculpt bioinspired technologies with tunable charge transfer, energy conversion, and photocatalytic properties. Yet, despite decades of research, the physical mechanisms that determine their electronic spectra remain elusive. Even for metal-free porphyrins, no consensus exists on the origin of the splitting of their $Q$-bands, an energetic region critical in photosynthesis. We leverage our recent statistical treatment of molecular motions in the condensed phase to predict their linear absorption spectra. By bridging exact quantum-dynamical expressions with atomistic simulations, our theory is the first to capture the spectra of representative porphyrins in solution: porphine, tetraphenylporphyrin, and tetraphenylporpholactone. Our work reveals that $Q$-band splitting arises from extreme timescale separation of nuclear motions that turn symmetry-forbidden transitions bright. We exploit these insights to propose and confirm how chemical modifications tune the optical properties of porphyrins, demonstrating the potential for developing design principles to tailor their optoelectronic behavior. 
\end{abstract}

\date{\today}
\maketitle

Porphyrins are ubiquitous in biology and chemistry and play crucial roles in, for example, oxygen transport and storage in blood, charge transfer in cellular respiration, and photosynthesis in plants\cite{mercer1984photochemistry, Porphyrinfundamentals2022}. Their synthetic tunability\cite{tanaka2015conjugated} make porphyrins ideal in advancing bioinspired charge and energy transport and storage technologies\cite{mathew2014dye, swierk2015metal, zhang2023h2o2, kumagi2014}, especially as they offer a convenient chemical scaffold that can be functionalized with substituents\cite{ning2020split, roy2022synthetic}, paired with transition metals\cite{moisanu2023crystalline}, or wreathed together in polymer chains\cite{gotfredsen2022bending,kosugi2023iron, magnera2023porphene, chen2024porphyrin} to tailor and exploit their optoelectronic properties. Numerous studies have turned to sophisticated multidimensional spectroscopies to disentangle the palette of spectral features generated by porphyrin and porphyrin-like systems\cite{Niedringhaus2018, policht2018characterization,  song2019vibronic, ArsenaultQxQy2020, Jana2023, zhao2023quantum, petropoulos2024vibronic}. Yet, predicting and interpreting even the linear electronic spectra of small porphyrin monomers remains challenging as their photophysics continues to elude a conclusive microscopic interpretation. 

In photosynthesis, energy transfer between carotenoids and porphyrin-based chlorophyll $Q$ bands facilitate photoprotection\cite{Ma2003, Holt2005, Ruban2007, polívka2010molecular, Bode2009, Son2020}. These $Q$ bands in free-base porphyrins present four peaks corresponding to two distinct electronic transitions, in contrast to their metalloporphyrin counterparts which present only two peaks. The widely accepted group-theoretic description of porphyrin spectra\cite{Gouterman_FourOrbital_1963} interprets these transitions as having mutually orthogonal transition dipoles in the plane of the tetrapyrrole ring: the lower-energy half of the $Q$ bands ($\rm S_0 \to S_1$) denoted as the $Q_x$ band and the higher-energy half ($\rm S_0 \to S_2$) denoted as the $Q_y$ band, with transition dipoles aligned parallel and orthogonal to the central proton axis, respectively. Yet, even the pioneering work of Gouterman\cite{gouterman1961spectra, Gouterman_FourOrbital_1963}, which explained the competition of intensities and center frequencies of the Soret and $Q$ bands in metalloporphyrins, does not explain why the $Q$ band peaks split from two to four upon removal of the metal center. Crucially, the microscopic mechanism underlying this spectral splitting continues to spark disagreement\cite{gouterman1961spectra, makriflemmingTPPHT, baskin2002ultrafast, schalk2008near, he2012vibronic, kullmann2012ultrafast}. Indeed, this doubling (four peaks instead of two) of the $Q$ bands has been variously ascribed to a Condon-allowed vibronic progression\cite{gouterman1961spectra, baskin2002ultrafast}, where the transition dipole is assumed to be insensitive to nuclear motions, or non-Condon effects via symmetry-breaking nuclear motions that tune the brightness, or selection rules, of transitions\cite{makriflemmingTPPHT, neville2024}. Given this controversy, it is not surprising that this critical problem in porphyrin photophysics has been the subject of intense theoretical focus.

Yet, state-of-the-art theoretical spectroscopy tools can only tackle isolated aspects of this question. For example, one can capture non-Condon effects under the microscopic harmonic approximation using the Franck-Condon Herzberg-Teller (FCHT) method\cite{Santoro_2008} but only in the gas phase and for fairly small and rigid molecules \cite{Zuehlsdorff2019}; address non-Condon fluctuations with atomistic anharmonicities in the condensed phase but only in the static limit (Ensemble)\cite{bergsma1984electronic}; or treat the dynamics and atomistic anharmonicities in the condensed phase but only in the Condon limit (2nd order cumulant)\cite{mukamel1985fluorescence, Zuehlsdorff2019}. Thus, a more general approach is required.

\begin{figure}[t]
    \centering
    \includegraphics[width=1\columnwidth]{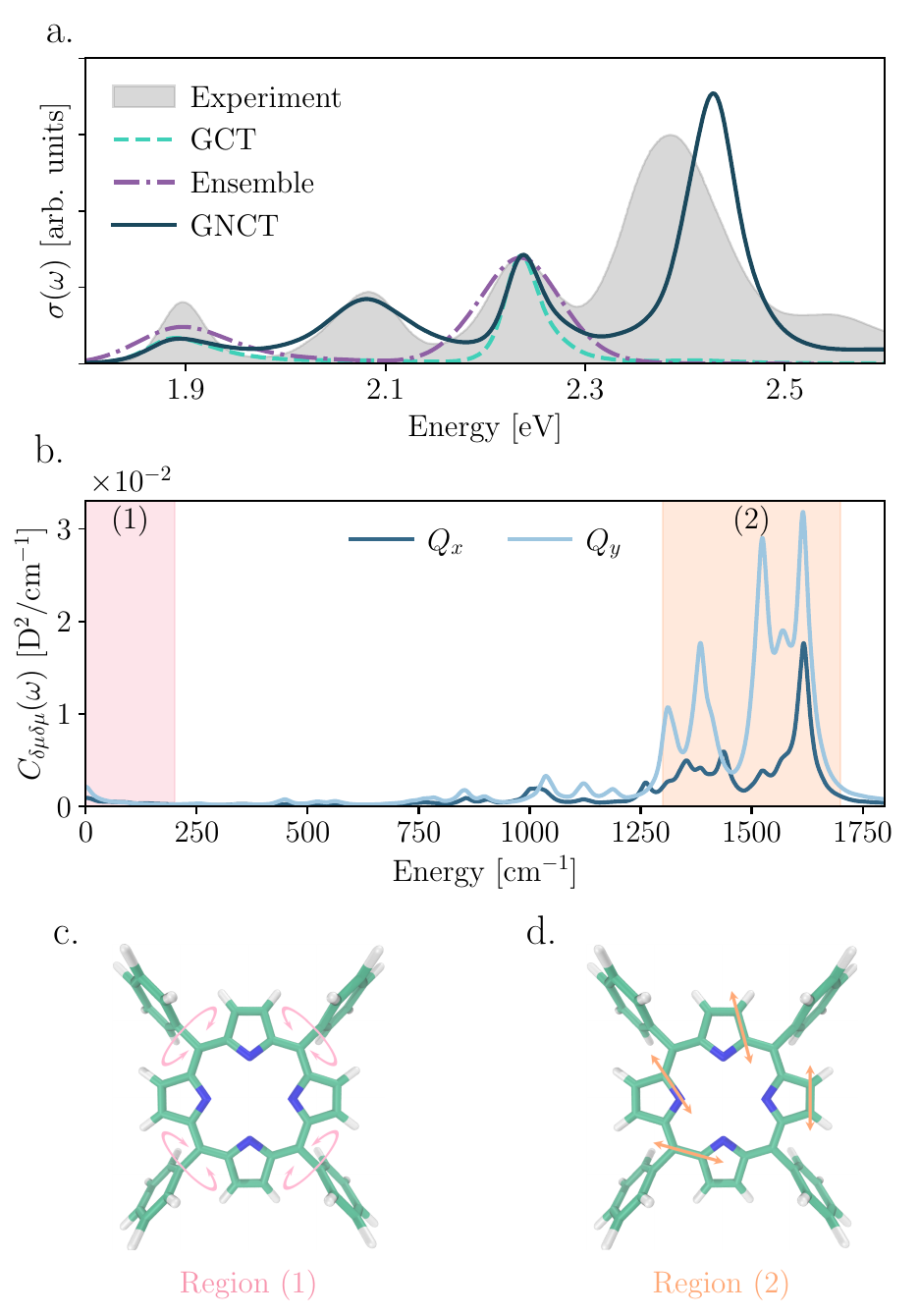}
    \caption{(a) The experimental $Q$-band absorption spectrum of free-base TPP (grey) overlaid by simulated spectra: GNCT (navy), GCT (teal) and ensemble (purple). Here, all simulated spectra are normalized to the third peak in the experimental spectrum. (b) The $Q_x$ and $Q_y$ transition dipole spectral densities of TPP where we have highlighted two regions corresponding to slow and fast vibrations in TPP: Region (1) arising from very slow torsions of the pendant phenyls (c) and, region (2) arising from fast inner-ring vibrations (d).}
    \label{fig:fig1-tpp-methods}
\end{figure}

This impasse has led to seemingly contradictory explanations of the physical origin of the $Q$ bands. One of the dominant interpretations\cite{baskin2002ultrafast} assigns the first two peaks (1.9-2.2 eV) of the tetraphenylporphyrin (TPP) $Q$ band (see, for example, the experimental spectrum of TPP in $\rm CS_2$ in Fig.~\ref{fig:fig1-tpp-methods}) as the 0-0 and 0-1 features in the vibronic progression of the S$_1$ ($Q_x$) electronic transition, and the latter two (2.2-2.5 eV) as the 0-0 and 0-1 vibronic progression of the S$_2$ ($Q_y$) transition. However, recent studies have used FCHT-type arguments to assign the $Q$ bands of porphine to mixed Condon-allowed (first and third peaks) and non-Condon-enabled electronic transitions (second and fourth)\cite{Santoro_2008, makriflemmingTPPHT}. Yet, these interpretations do not provide a fully predictive theory of the lineshape that unambiguously captures the splitting and broadening of spectral peaks. Hence, it is crucial to employ a theoretical framework that can capture \textit{both} dynamical vibronic progressions and the non-Condon fluctuations that tune transition brightness as molecules vibrate and jostle in solution.

We bridge this theoretical gap by leveraging our recent theory of non-Condon fluctuations that offers a quantum dynamical means to process the output of classical molecular dynamics (MD) simulations and electronic structure theory to predict linear optical spectra in the condensed phase \cite{wiethornGNCTtheory2023}. Our theory predicts the optical response of solvated chromophores while faithfully incorporating their anharmonic atomistic interactions, arbitrarily strong non-Condon effects (nuclear motion-induced variations of the transition dipoles), and significant environmental coupling. Our expressions encode the chemical details that determine optical spectra through spectral densities that quantify how nuclear motions modulate the energy gap, the transition dipole, and their correlations. Crucially, one can \textit{directly} calculate these spectral densities from atomistic simulations of a material, offering a direct route to identifying and interpreting the optical response. This ultimately allows us to resolve the long-standing controversy over the physical origin of the $Q$ band splitting of porphyrins by quantifying the influence of the following processes: 
\vspace{-3pt}
\begin{enumerate}
    \item Vibronic features on Condon-allowed transitions.
    \vspace{-5pt}
    \item Static and dynamic non-Condon fluctuations.
    \vspace{-5pt}
    \item Energy transfer between $Q_x$ and $Q_y$ bands.
    \vspace{-5pt}
\end{enumerate} 
Our work is the first to accurately reproduce and explain the spectral splitting and intensity distribution of the $Q$ bands of representative porphyrins: porphine, TPP, and tetraphenylporpholactone (TPPL). Our analysis reveals that the $Q$-band splitting arises from extreme timescale separation between the slow torsions of the pendant phenyls (in TPP and TPPL) and collective solvent motions and the fast tetrapyrrole ring vibrations that modulate the transition dipole, making this splitting a primarily non-Condon feature. We further introduce a perturbative treatment of excited-state energy transfer to confirm that this splitting is robust to energy flow across the $Q$ bands. 

Our theory of non-Condon spectroscopy exploits the idea that the macroscopic number of nuclear motions that modulate the energy gap $U(\hat{\mathbf{q}})$ and transition dipole $\boldsymbol{\mu}(\hat{\mathbf{q}})$ in the condensed phase are likely to cause Gaussian fluctuations. This enables us to write an effective Gaussian model for which we have derived exact closed-form expressions for the optical response function \cite{wiethornGNCTtheory2023} that can be parameterized directly from MD simulations (see Methods Section \ref{Methods_Sec:Theory-GNCT-nature}). In contrast to previous works, our Gaussian non-Condon theory (GNCT) does not resort to fitting parameters that restrict its predictive power and can lead to ill-converging spectra when strongly anharmonic nuclear motions render harmonic approximations inapplicable (see FCHT prediction in Fig.~\ref{fig:fig5-Jsd-duscinsky-ginfty}(c). In addition, it enables us to isolate contributions to the absorption spectrum, $\sigma (\omega)$, arising from transition dipole fluctuations that are static and dynamic relative to electronic dephasing times,
\begin{equation}\label{nature-main-eq:GNCT-total-spectrum}
   \sigma^{}_{\rm GNCT} (\omega)  \! =  \! \alpha( \omega ) \! \left[ \sigma^{\boldsymbol{\mu}{\rm -stat.}}_{\rm GNCT} (\omega) \! + \! \sigma^{\mu{\rm-dyn.}}_{\rm GNCT} (\omega)  \right] \!. 
\end{equation}
Here, $\alpha(\omega)$ is the linear absorption coefficient\cite{Mukamel-book} centered at $\omega=0$ where as $\sigma^{\boldsymbol{\mu}{\rm -stat.}}_{\rm GNCT} (\omega)$ and $\sigma^{\mu{\rm-dyn.}}_{\rm GNCT} (\omega)$ are shifted by the average transition energy $\langle \omega_{eg} \rangle$ between the ground ($g$) and excited ($e$) state. Below, we describe the physical meaning of the two components in Eq.~\eqref{nature-main-eq:GNCT-total-spectrum} within the pure-dephasing limit (i.e., in the absence of energy transfer), and discuss analogous expressions we derive and employ to capture the influence of excited-state energy flow on the spectrum in Sec.~IV of the supplementary information (SI). 

Our GNCT predicts that the average transition dipole, $\langle \hat{\boldsymbol{\mu}}_{ge} \rangle$, determines the magnitude of the static non-Condon contribution to the spectrum,
\begin{equation}\label{nature-main-eq:NC-static-spectrum}
    \sigma^{\mu{\rm -stat.}}_{\rm GNCT} (\omega) = | \langle \hat{\boldsymbol{\mu}}^{}_{ge} \rangle|^2 \,  \sigma^{}_{\delta U \delta U} (\omega).
\end{equation}
Here, 
$\langle \hat{\boldsymbol{\mu}}_{ge} \rangle $ arises from the average effect of nuclear motions that break the symmetry of a molecule. When $\sigma^{\mu{\rm -stat.}}_{\rm GNCT} (\omega)$ dominates the spectrum, it captures non-Condon disorder while accounting for the energy gap dynamics, which encode vibronic progressions. When $ \langle \hat{\boldsymbol{\mu}}_{ge} \rangle $ is insensitive to nuclear configurations, it can be replaced by the transition dipole at any nuclear configuration, recovering the celebrated second-order cumulant within the Condon approximation, $| \hat{\boldsymbol{\mu}}^{}_{ge}|^2 \sigma_{\delta U \delta U}(\omega)$\cite{Zuehlsdorff2019, mukamel1985fluorescence} (see Methods~A for the full expression). The energy gap spectral density, $C_{\delta U \delta U} (\omega) = \left[ 1 + \coth(\omega/k_{\rm B} T) \right] C_{\delta U \delta U}'' (\omega)$\cite{Mukamel-book}, encodes the dynamical energy gap fluctuations ($\delta U (\hat{\mathbf{q}})$) that control the lineshape of $\sigma_{\delta U \delta U}(\omega)$ and its ability to exhibit distinct vibronic progressions or its related asymmetry.

Our GNCT quantifies how dynamic non-Condon fluctuations impact the spectrum via $s_{\delta\boldsymbol{\mu} \delta\boldsymbol{\mu}} (\omega)$ and $\mathbf{s}_{\delta U \delta\boldsymbol{\mu}} (\omega)$, 
\begin{equation}\label{nature-main-eq:NC-dynamic-spectrum}
\begin{split}
\sigma^{\boldsymbol{\mu}{\rm -dyn.}}_{\rm GNCT} (\omega) =& \big[ \big(s_{\delta \boldsymbol{\mu} \delta \boldsymbol{\mu}} \! + \! 2 \langle \hat{\boldsymbol{\mu}}_{ge} \rangle \cdot  \mathbf{s}_{\delta U \delta \boldsymbol{\mu}} \big) \! \ast \! \sigma_{\delta U \delta U} \big] \! (\omega) \\& + \big[ \mathbf{s}_{\delta U \delta \boldsymbol{\mu}} \! \ast\! \mathbf{s}_{\delta U \delta \boldsymbol{\mu}} \! \ast \! \sigma_{\delta U \delta U} \big] \!(\omega).
\end{split}
\end{equation}
Just as $C_{\delta U \delta U} (\omega)$ encodes how energy gap fluctuations control the prominence of vibronic progressions in $\sigma^{}_{\delta U \delta U} (\omega)$, here, two new spectral densities, $C_{\delta\boldsymbol{\mu} \delta\boldsymbol{\mu}}(\omega)$ and $\mathbf{C}_{\delta U \delta\boldsymbol{\mu}}(\omega)$, encode how nuclear motions cause dynamic transition dipole fluctuations ($\delta \boldsymbol{\mu}(\hat{\mathbf{q}};t)$) and determine $s_{\delta\boldsymbol{\mu} \delta\boldsymbol{\mu}} (\omega)$ and $\mathbf{s}_{\delta U \delta\boldsymbol{\mu}} (\omega)$. $C_{\delta\boldsymbol{\mu} \delta\boldsymbol{\mu}}(\omega)$ quantifies how nuclear motions modulate the brightness of transitions while $\mathbf{C}_{\delta U \delta\boldsymbol{\mu}} (\omega)$ measures how nuclear motions jointly tune the energy gap and transition brightness. Together, Eqs.~\eqref{nature-main-eq:NC-static-spectrum} and \eqref{nature-main-eq:NC-dynamic-spectrum} determine the optical spectrum arising from static and dynamic transition dipole modulations. Our GNCT thus reveals that dynamic non-Condon effects arise from the convolution (denoted by $[f \ast g](\omega)$) of the energy gap lineshape, $\sigma_{\delta U \delta U}(\omega)$, with the contributions arising from the transition dipole fluctuations.

Because our theory provides a transparent protocol to parameterize Eqs.~\eqref{nature-main-eq:NC-static-spectrum} and \eqref{nature-main-eq:NC-dynamic-spectrum} using atomistic simulations, we can assign spectral signals to specific nuclear motions and assess their effect on both energy gap and transition dipole fluctuations. We now show how our GNCT, unlike previous theories, recapitulates the condensed phase spectra of porphyrins and uncovers the mechanism leading to the splitting of their $Q$ bands as a function of their chemical structure.

We first ask if the observed peak splitting arises purely from Condon-allowed vibronic progressions. The second-order cumulant method in the Condon limit (to which we refer henceforth as ``GCT" for Gaussian Condon theory) successfully captures vibronic progressions\cite{mukamel1985fluorescence,loco2018modeling,Zuehlsdorff2019}, and is analogous to $\sigma_{\rm GNCT}^{\mu{\rm-stat.}} (\omega)$ in our GNCT (Eq.~\eqref{nature-main-eq:NC-static-spectrum}) in the Condon limit (i.e., $\langle \hat{\boldsymbol{\mu}}_{ge} \rangle \to \hat{\boldsymbol{\mu}}_{ge}$). Figure~\ref{fig:fig1-tpp-methods} shows that the predicted GCT spectrum (teal) only recovers the first ($Q_x$) and third ($Q_y$) peaks, revealing that the $Q$-band splitting does not arise from a Condon-allowed vibronic progression and non-Condon effects must be considered.

\begin{figure*}
    \includegraphics[width=1\textwidth]{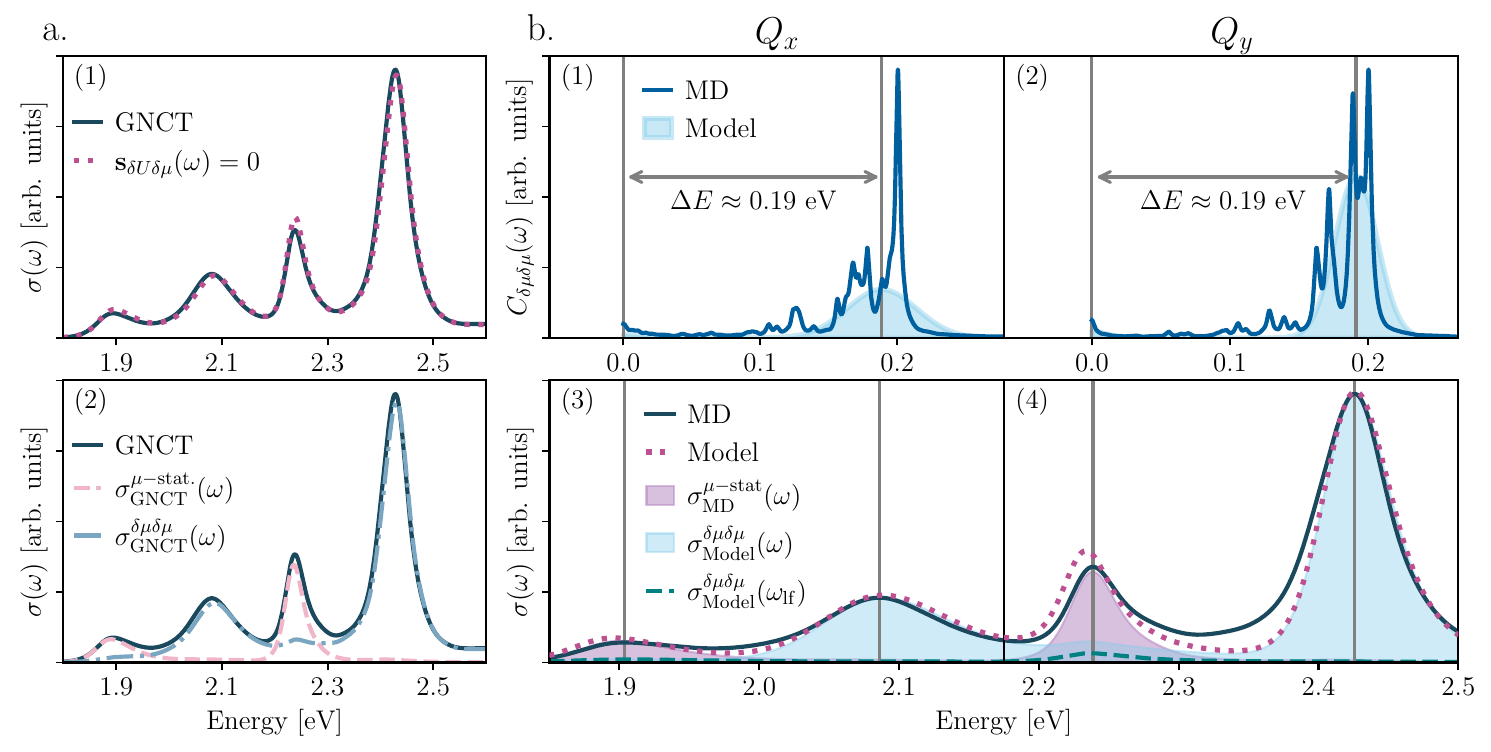}
    \caption{ Investigation of how each component of our GNCT contributes to the splitting of the TPP $Q$ bands. Panel (a) uncovers that the shared nuclear motion-induced modulations of the excitation energies and transition dipoles do not cause peak splitting (a1) and the emergence of four peaks is driven entirely by $\sigma_{\rm GNCT}^{\boldsymbol{\mu}-\rm stat.} (\omega)$ and $\sigma_{\rm GNCT}^{\delta \boldsymbol{\mu} \delta \boldsymbol{\mu}} (\omega) = s_{\delta \boldsymbol{\mu} \delta \boldsymbol{\mu}}(\omega) \ast \sigma_{\delta U \delta U} (\omega)$ (a2). Panel b compares our $C^{\rm model}_{\delta \boldsymbol{\mu} \delta \boldsymbol{\mu}} (\omega)$ prediction against the MD-based results. Panels (b1) and (b2) show the model spectral densities (shaded) that we substituted for the atomistic spectral densities (line) for $Q_x$ and $Q_y$ bands, respectively. The spectra in panels (b3) and (b4) show how our model-derived result (dashed) agree well with that obtained from MD (solid). The shaded regions unveil that the first and third peaks arise from $\sigma_{\rm GNCT}^{\mu-\rm stat.} (\omega)$ (pink) whereas the second and fourth peaks arise largely from the dynamical $\sigma_{\rm GNCT}^{\delta \boldsymbol{\mu} \delta \boldsymbol{\mu}} (\omega)$ (blue) indicating that a strong separation of timescales between static and dynamic transition dipole fluctuations cause the $Q_x$ and $Q_y$ bands of TPP to split.}
    \label{fig2:tpp_md_and_model_decomp}
\end{figure*}

Since Condon-allowed transitions do not cause the peak splitting, we turn to \textit{static} non-Condon effects. The ensemble method\cite{bergsma1984electronic}, widely used to predict condensed phase spectra\cite{zuehlsdorff2021ARPC}, treats non-Condon fluctuations under the approximation that nuclear motions are slower than electronic dephasing times, thereby capturing the static limit of \textit{both} non-Condon and energy gap fluctuations. However, similar to the GCT, Fig.~\ref{fig:fig1-tpp-methods} shows that the ensemble method (purple) can capture only the first ($Q_x$) and third ($Q_y$) peaks. Hence, the splitting does not arise solely from static non-Condon effects. 

To incorporate both static \textit{and} dynamic non-Condon fluctuations, we turn to our GNCT. Figure~\ref{fig:fig1-tpp-methods} illustrates that our GNCT (dark green) is the only method that correctly reproduces the four peaks of the TPP $Q$ bands, their relative intensities, and even their linewidths. Thus, peak splitting in the TPP $Q$ bands appears to arise from \textit{dynamic non-Condon fluctuations}.

We are now poised to interrogate the \textit{microscopic mechanism} leading to $Q$ band splitting. First, Fig.~\ref{fig2:tpp_md_and_model_decomp}(a.1) shows that excluding $\mathbf{s}_{\delta U \delta\boldsymbol{\mu}} (\omega)$ (dotted) causes only negligible changes to the intensity distribution of the spectrum (see also SI Fig.~S1), demonstrating that the joint modulation of transition dipoles and energy gaps by nuclear motions does not participate in splitting the TPP $Q$ bands. Further, Fig.~\ref{fig2:tpp_md_and_model_decomp}(a.2) shows that zeroing the purely dynamical non-Condon contribution, $\sigma_{\rm GNCT}^{\delta\boldsymbol{\mu} \delta\boldsymbol{\mu}}(\omega)~\equiv~[s_{\delta \boldsymbol{\mu} \delta \boldsymbol{\mu}}\ast \sigma_{\delta U \delta U}](\omega)$, causes the second and fourth peaks in the TPP spectrum to disappear (pink-dashed). Since this is tantamount to neglecting the full dynamical non-Condon contribution, $\sigma^{\mu{\rm-dyn.}}_{\rm GNCT} (\omega)$, one can assign the second ($Q_x$) and fourth ($Q_y$) peaks in TPP to purely dynamical non-Condon effects. In contrast, zeroing the static non-Condon component, $\sigma^{\mu{\rm-stat.}}_{\rm GNCT} (\omega)$, decimates the first ($Q_x$) and third ($Q_y$) peaks (blue-dashed), revealing these as largely static non-Condon contributions. This suggests that peak splitting in TPP arises from a static-dynamic divide in the nuclear motion-induced modulation of the selection rules. However, the persistence of small components in the first and third peaks upon neglecting $\sigma^{\mu{\rm-stat.}}_{\rm GNCT} (\omega)$ motivates a more microscopic interrogation of $\sigma^{\delta\boldsymbol{\mu} \delta\boldsymbol{\mu}}_{\rm GNCT}(\omega)$.

To uncover the origin of the persistent splitting in the surviving dynamical non-Condon contribution to the GNCT lineshape, $\sigma_{\rm GNCT}^{\delta\boldsymbol{\mu} \delta\boldsymbol{\mu}}(\omega)$, we analyze the transition dipole spectral density that determines its behavior, $C_{\delta\boldsymbol{\mu} \delta\boldsymbol{\mu}}(\omega)$ (see Fig.~\ref{fig:fig1-tpp-methods}(b)). $C^{Q_x}_{\delta\boldsymbol{\mu} \delta\boldsymbol{\mu}}(\omega)$ and $C^{Q_y}_{\delta\boldsymbol{\mu} \delta\boldsymbol{\mu}}(\omega)$ display complex structure with a small contribution at low frequency ($<200$ cm$^{-1}$) and a region of considerable spectral weight centered around $\sim 1500 $ $ \rm cm^{-1}$. Restricted normal mode analyses in the gas phase enable us to assign the high-frequency region of the spectral density (region 2 in Fig.~\ref{fig:fig1-tpp-methods}(b)) to inner-ring C-C and C-N vibrations (see SI Sec.~II). The low-frequency contribution below $<200$ cm$^{-1}$ arises from collective solvent motions in solution and torsional motions of the phenyl substituents with timescales $\sim 33$ cm$^{-1}$ (see SI Sec.~III). But to what extent does the structure of these spectral densities matter and how does this apparent timescale separation in nuclear motions contribute to the splitting of the $Q$-bands? 

To address the first question, we replace the MD-derived transition dipole spectral densities ($C_{\delta \boldsymbol{\mu} \delta \boldsymbol{\mu}}(\omega)$) with model counterparts composed of low- and high-frequency contributions, $C^{\rm model}_{\delta\boldsymbol{\mu} \delta\boldsymbol{\mu}}(\omega) = C^{\rm model}_{\delta\boldsymbol{\mu} \delta\boldsymbol{\mu}}(\omega_{\rm lf}) + C^{\rm model}_{\delta\boldsymbol{\mu} \delta\boldsymbol{\mu}}(\omega_{\rm hf})$. For $C^{\rm model}_{\delta\boldsymbol{\mu} \delta\boldsymbol{\mu}}(\omega_{\rm lf})$, we employ an often adopted Debye form,\cite{Weiss-book} and subsume the high-frequency intramolecular vibrations into a Gaussian envelope for $C^{\rm model}_{\delta\boldsymbol{\mu} \delta\boldsymbol{\mu}}(\omega_{\rm lf})$ (see Methods Sec.~\ref{Methods_Sec:Model-system-spectral-densities-GNCT-nature}). Figure~\ref{fig2:tpp_md_and_model_decomp}(b.1)-(b.2) illustrate the MD-based (line) and our simplified (shaded) $C_{\delta\boldsymbol{\mu} \delta\boldsymbol{\mu}}(\omega)$ for the $Q_x$ and $Q_y$ bands, with their corresponding absorption spectra in (b.3)-(b.4). Our simplified spectral densities faithfully reproduce the splitting of the $Q$-bands (dashed-peach), agreeing semi-quantitatively with the MD-based spectrum (solid), indicating that the detailed structure of the MD-derived spectral density is not essential to recover the TPP $Q$ band splitting and validates our use of these models in our investigation of timescale separation.

We now assess the influence of timescale-separated fluctuations in $\sigma^{\rm model}_{\delta \boldsymbol{\mu} \delta \boldsymbol{\mu}} (\omega)$. By turning off the high-frequency Gaussian envelopes in $C^{\rm model}_{\delta\boldsymbol{\mu} \delta\boldsymbol{\mu}}(\omega)$, one observes that the second ($Q_x$) and fourth ($Q_y$) peaks vanish in Fig.~\ref{fig2:tpp_md_and_model_decomp}(b.3) and (b.4), leaving only the small contributions to the first ($Q_x$) and third ($Q_y$) peaks (green-dashed), which we had largely assigned as static non-Condon contributions. What is more, the energy difference between split $Q$ band peaks (0.19 eV) in Fig.~\ref{fig2:tpp_md_and_model_decomp}(b.3)-(b.4) matches the energetic split between the low- versus high-frequency components of the simplified non-Condon spectral densities in Fig.~\ref{fig2:tpp_md_and_model_decomp}(b.1)-(b.2). A consistent picture emerges: the first ($Q_x$) and third ($Q_y$) peaks arise from static ($\langle \hat{\boldsymbol{\mu}}_{ge} \rangle$) and slow (collective solvent motions and phenyl torsions) non-Condon contributions, while second ($Q_x$) and fourth ($Q_y$) peaks arise from the fast modulation of the transition dipoles arising from intra-tetrapyrrole ring vibrations, revealing that \textit{the extreme timescale separation of the nuclear motions that modulate the transition dipole, and thus the optical selection rules, offers the mechanism that splits the TPP $Q$ bands.}

While our GNCT addresses the pure dephasing limit where no energy transfer occurs between excited states, TPP is known to undergo internal conversion as energy flows from the $Q_y$ to the $Q_x$ band\cite{baskin2002ultrafast}. Hence, we developed a framework to quantify the impact of $Q_y \rightarrow Q_x$ energy flow on the spectrum (see SI Sec.~IV) by treating the driver of energy flow as a perturbation. We tailor an excited state transformation\cite{Subotnik2017} to move into the frame where states have spatially orthogonal transition dipoles, consistent with the traditional interpretation of porphyrin photophysics (see SI Sec.~IVB). We find that including energy flow across the $Q$ bands has a marginal effect on the computed spectrum (Fig.~S10), indicating that our timescale separation mechanism for $Q$ band splitting remains valid even in the presence of internal conversion. 

Besides offering mechanistic insights, our GNCT yields design principles to tune optoelectronic properties. Specifically, the spectral splitting mechanism in porphyrins suggests that one can balance the contributions of $\sigma^{\mu\rm-stat.}_{\rm GNCT} (\omega)$ and $\sigma^{\mu\rm-dyn.}_{\rm GNCT} (\omega)$ via simple chemical changes. For instance, we hypothesize that one can suppress the first and third peaks of the porphyrin $Q$ bands by removing the meso-phenyls or replacing them with bulky chemical moieties. In contrast, asymmetrically modifying the tetrapyrrole ring by hydrating one of the pyrroles to form a chlorin or substituting a functional group into the ring can be expected to: (1) change the intensity distribution across the $Q$ band by modulating $|\langle \boldsymbol{\hat{\mu}}_{ge} \rangle|^2$ via steric or electron conjugation effects that tune the rotation of the pendant phenyls; and (2) modify the individual $Q_x$/$Q_y$ lineshapes by tuning the energy gap spectral density $C_{\delta U \delta U} (\omega)$ which determines the prominence of vibronic progressions in $\sigma_{\delta U \delta U}(\omega)$. To assess the validity of our design hypotheses, we turn to porphine and TPPL. These molecules offer an ideal testing ground since replacing the pendant phenyls in TPP with hydrogens yields porphine, whereas TPPL retains TPP's phenyls but replaces an external ($\beta$-periphery) $\rm C_2H_2$ on one pyrrole orthogonal ($y$) to the proton axis ($x$) with a lactone. 

\begin{figure}[t]
    \includegraphics[width=1\columnwidth]{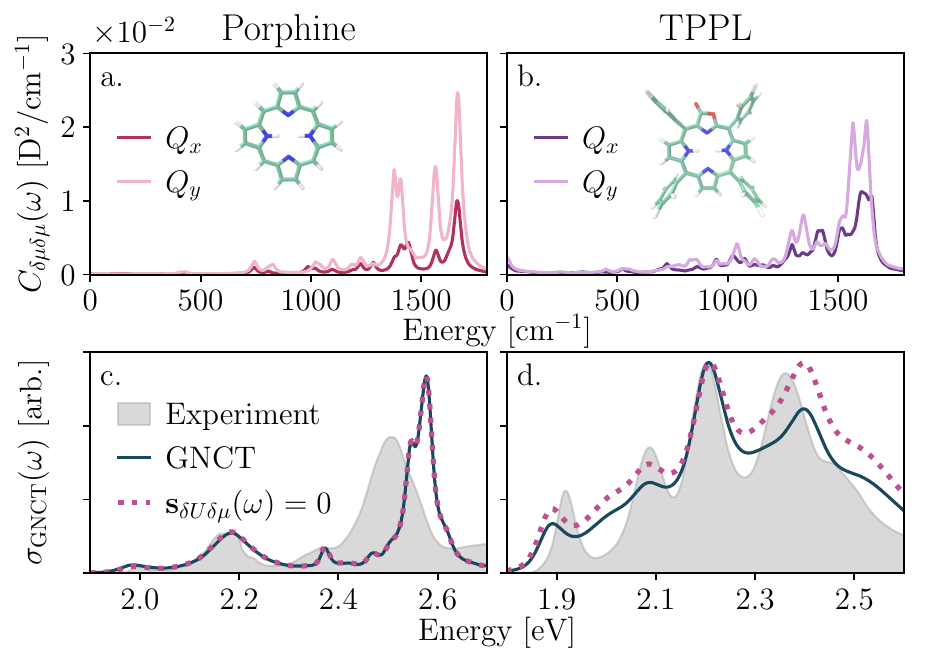}
    \caption{Transition dipole spectral densities (top) and $Q$ bands (bottom) of porphine and TPPL. In panel (a) one observes that removing the meso-phenyls from TPP causes the low-frequency region of $C^{Q_x}_{\delta\boldsymbol{\mu} \delta\boldsymbol{\mu}}(\omega)$ and $C^{Q_y}_{\delta\boldsymbol{\mu} \delta\boldsymbol{\mu}}(\omega)$ of porphine to nearly vanish. In contrast in going to TPPL in panel (b) the lactone substitution enhances the low-frequency region of $C^{Q_x}_{\delta\boldsymbol{\mu} \delta\boldsymbol{\mu}}(\omega)$ and $C^{Q_y}_{\delta\boldsymbol{\mu} \delta\boldsymbol{\mu}}(\omega)$. Panels (c) and (d) show our GNCT predicted spectra (solid) against experiment (shaded) in addition to the predicted spectra upon setting $\mathbf{s}_{\delta U \delta \boldsymbol{\mu}}(\omega) = 0$ (dashed) illustrating that, just as in TPP, the porphine and TPPL the $Q$ bands can be attributed to the relative strength between static and dynamic transition dipole fluctuations.}
    \label{fig:fig4-pophine-tppl-adiabat-dipole-spectral-density-and-spectra}
\end{figure}

We begin by leveraging the transition dipole spectral densities, $C_{\delta\boldsymbol{\mu}\delta\boldsymbol{\mu}} (\omega)$, to analyze the effect of the chemical modifications required to transform TPP to porphine and TPPL. We employ the non-Condon factor\cite{wiethornGNCTtheory2023}, $\varphi^\mu = \frac{C''_{\delta\boldsymbol{\mu}\delta\boldsymbol{\mu}} (\omega) - |\langle \hat{\boldsymbol{\mu}}_{ge} \rangle|^2 }{C''_{\delta\boldsymbol{\mu}\delta\boldsymbol{\mu}} (\omega) + |\langle \hat{\boldsymbol{\mu}}_{ge} \rangle|^2}$, to quantify how these changes tune the balance between static and dynamic non-Condon effects, with an increase (decrease) in $\varphi^{\mu}$ indicating that dynamical (static) non-Condon effects become more significant. Figure~\ref{fig:fig4-pophine-tppl-adiabat-dipole-spectral-density-and-spectra} shows $C^{Q_x}_{\delta\boldsymbol{\mu} \delta\boldsymbol{\mu}}(\omega)$ and $C^{Q_y}_{\delta\boldsymbol{\mu} \delta\boldsymbol{\mu}}(\omega)$ for porphine (a) and TPPL (b). In porphine, the absence of phenyl torsions that drive low-frequency transition dipole fluctuations in TPP leads to spectral densities with no low-frequency weight, and an increase in $\varphi_{Q_x}^{\mu}$ and $\varphi_{Q_y}^{\mu}$ by factors of 1.51 and 1.65, respectively, indicating that the optoelectronic properties of porphine depend more strongly on dynamic non-Condon fluctuations. 
In TPPL, these spectral densities exhibit significant low-frequency spectral weight from the phenyl torsions, albeit with greater intensity than in TPP. In contrast to TPP, one also observes enhanced intensity centering at 1000 $\rm cm^{-1}$ emerging from dynamic modulations of the transition dipoles by the C-O stretch. Moreover, the enhanced low-frequency motions lead to a reduction in $\varphi_{Q_x}^{\mu}$ and $\varphi_{Q_y}^{\mu}$ by factors of 2.42 and 2.79, respectively, indicating that static non-Condon effects are increasingly significant to the optoelectronic properties of TPPL as compared to TPP.

\begin{figure*}
    \includegraphics[width=1\textwidth]{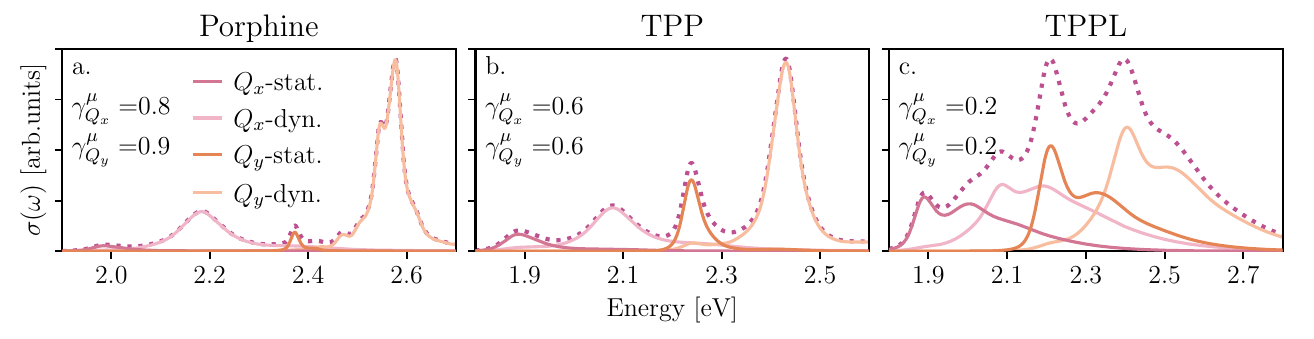}
    \caption{Quantifying the relative contributions from $\sigma^{\mu-\rm stat.}_{\rm GNCT} (\omega)$ (red-$Q_x$; orange-$Q_y$) and $\sigma^{\mu-\rm dyn.}_{\rm GNCT} (\omega)$ (pink-$Q_x$; peach-$Q_y$) to the $Q_x$ and $Q_y$ bands of porphine, TPP and TPPL, as a function of the non-Condon factor $\varphi^{\mu}$. By scanning across each pannel, one observes a gradual decrease in $\varphi^{\mu}$ as the contribution from $\sigma^{\mu-\rm stat.}_{\rm GNCT} (\omega)$ increases.}
    \label{fig:fig4-porpyrin-band-decomposition-static-and-dynamic}
\end{figure*}
We now turn to the optical spectra in Fig.~\ref{fig:fig4-pophine-tppl-adiabat-dipole-spectral-density-and-spectra}(c) and (d). In porphine and TPPL (as in TPP), setting $\mathbf{s}_{\delta U \delta\boldsymbol{\mu}} (\omega) = 0$ (dotted) reveals that $s_{\delta\boldsymbol{\mu} \delta\boldsymbol{\mu}} (\omega)$ captures the dynamic contribution to the $Q$ band splitting (see also SI Sec.~I). Hence, we analyze the resulting spectra through the lens of static and dynamic non-Condon contributions. Figure \ref{fig:fig4-pophine-tppl-adiabat-dipole-spectral-density-and-spectra}(c) shows the porphine spectrum and confirms our hypothesis that removing the phenyl groups reduces both the low-frequency dynamical and static contributions arising from non-Condon fluctuations, leading to suppressed first and third peaks in the porphyrin $Q$ bands. TPPL is more complex. Figure~\ref{fig:fig4-pophine-tppl-adiabat-dipole-spectral-density-and-spectra}(d) reveals that the TPPL $Q$ bands contain three features that are absent in porphine and TPP: (1) spectral intensity is greatest at the third peak rather than the fourth; (2) a vibronic shoulder now emerges from the fourth peak and; (3) the energetic separation of $\sim919.5$ cm$^{-1}$ between second (dynamic-$Q_x$) and third (static-$Q_y$) peaks deviates from the former $\sim1500$ cm$^{-1}$ splitting between subsequent peaks. Two of these changes align with our hypotheses: the shift in spectral intensity across the $Q$ bands from the lactone-induced change to the static-dynamic balance of non-Condon effects, and the larger vibronic progression from the lactone-induced modification of the high-frequency component of the energy gap spectral density that shifts peak prominence. However, the modified splitting between the second and third peaks is surprising, motivating a closer inspection of our hypotheses.

Hypothesis (1) centers on the balance between $\sigma^{\mu-\rm stat.}_{\rm GNCT} (\omega)$ and $\sigma^{\mu-\rm dyn.}_{\rm GNCT} (\omega)$. Figure~\ref{fig:fig4-porpyrin-band-decomposition-static-and-dynamic} decomposes our GNCT prediction for the $Q_x$ and $Q_y$ bands of porphine, TPP, and TPPL, illustrating the relative contributions of $\sigma^{\mu-\rm stat.}_{\rm GNCT} (\omega)$ (red and orange) and $\sigma^{\delta \boldsymbol{\mu} \delta \boldsymbol{\mu} }_{\rm GNCT} (\omega)$ (pink and peach) to the individual bands (dotted) as a function of the non-Condon factors. This figure shows that the $\sigma^{\mu-\rm stat.}_{\rm GNCT} (\omega)$ contribution is negligible in porphine, becomes significant in TPP, and reaches commensurate values with the dynamic contribution in TPPL, thus aligning with our hypothesis that the phenyl groups continue to emphasize the first and third peaks relative to the second and fourth. Figure~\ref{fig:fig4-porpyrin-band-decomposition-static-and-dynamic} further reveals the true complexity of the TPPL spectrum as all four contributions overlap significantly, with each displaying a prominent lactone-induced vibronic progression. This is in stark contrast to the four unimodal peaks that lack energy gap-derived vibronic structure underlying the $Q$ bands of porphine and TPP. This observation also confirms hypothesis (2) as it reveals that the asymmetric lactone functionalization of the inner ring induces vibronic progressions that can redistribute spectral weight. 

Yet, it seems that a contradiction lingers in our analysis. Figure~\ref{fig:fig4-pophine-tppl-adiabat-dipole-spectral-density-and-spectra}(d) shows that the third peak in the experimental TPPL spectrum carries greater intensity than the fourth peak, which one may naively ascribe to a greater static non-Condon contribution in the $Q_y$ band based on our current analysis. However, Fig.~\ref{fig:fig4-porpyrin-band-decomposition-static-and-dynamic}(c) shows the opposite is true: the dynamical non-Condon $Q_y$ peak is predicted to be greater. This seeming contradiction is a result of considering the static and dynamic non-Condon peaks separately. Fortuitously, the peak separation in the vibronic progression of the dynamic non-Condon $Q_x$ feature (light pink solid line) is nearly identical to the peak separation between dynamic $Q_x$ ($\sim2.1$~eV) and static $Q_y$ ($\sim2.2$~eV) in the full spectrum. As a result, the overlapping dynamic $Q_x$ satellite peak (vibronic progression) supplies spectral weight to the static non-Condon $Q_y$ peak at $\sim2.2$~eV, causing it to appear to dominate over that of the dynamic $Q_y$ at $\sim2.42$~eV in the total spectrum. This analysis thus demonstrates how the interplay between Condon-allowed vibronic progressions and symmetry-forbidden transitions control the TPPL $Q$-band intensity distribution. Yet, despite our GNCT's success in predicting the relative peak heights, it appears to overestimate the prominence of a lactone-induced vibronic progression for the static $Q_x$ peak that is absent in the experimental spectrum (consider the $\sim 0.17$ eV gap between first and second peaks in Fig.~\ref{fig:fig4-pophine-tppl-adiabat-dipole-spectral-density-and-spectra}(d)). Since energy gap spectral densities encode vibronic progressions, an analysis of $C^{Q_x}_{\delta U \delta U}(\omega)$ ought to reveal why our GNCT predicts this spurious effect in the TPPL $Q_x$ band.

\begin{figure*}
    \centering
    \includegraphics[width=1.0\textwidth]{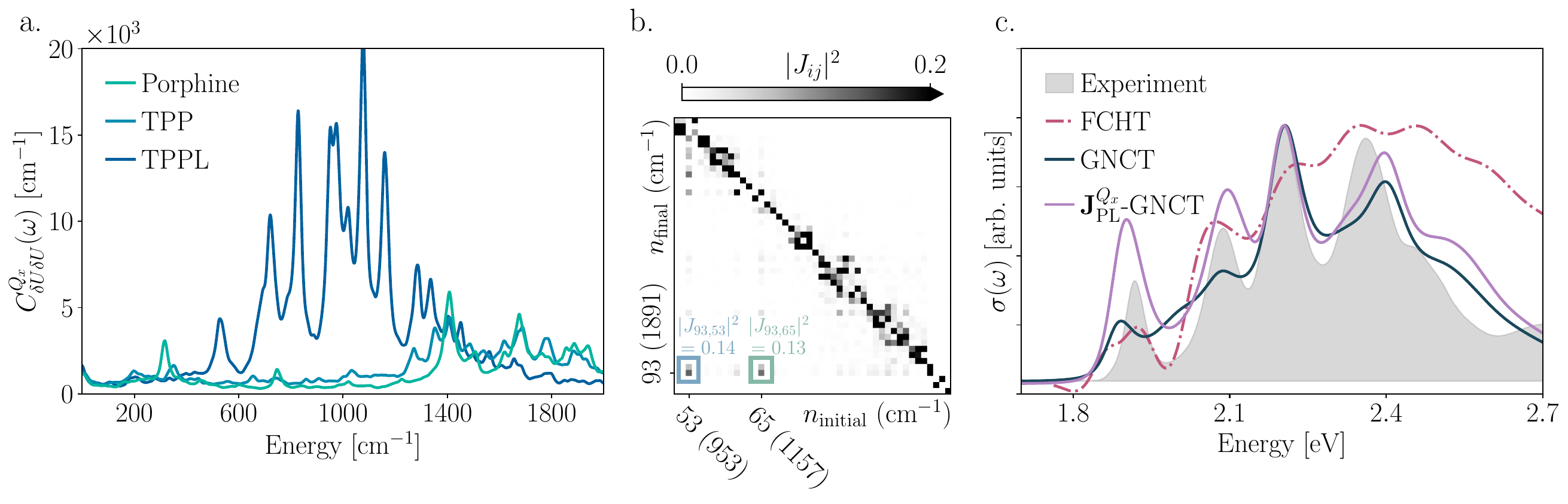}
    \caption{(a) The MD-derived energy gap spectral densities for the $Q_x$ band of porphine, TPP and TPPL showing that in the range of $600 - 1400$ cm$^{-1}$, TPPL displays significant spectral weight whereas TPP and prophine do not. (b) Subsection of Duschinsky matrix highlighting the prominence of mode mixing between the excited-state porphalactone carbonyl stretch occurring at 1891 cm$^{-1}$ and the ground-state in-plane stretching of the inner-ring C-O (953 cm$^{-1}$, blue) and an asymmetric lactone stretch (1157 cm$^{-1}$, green). (c) Comparison of predicted TPPL $Q$ bands against experiment showing that the FCHT yields unphysical negative absorption whereas our GNCT, and the generalized $\mathbf{J}_{\rm PL}^{Q_x}$-GNCT to account for non-Gaussian mode mixing, provide a physical and accurate absorption spectrum.}
    \label{fig:fig5-Jsd-duscinsky-ginfty}
\end{figure*}
Figure~\ref{fig:fig5-Jsd-duscinsky-ginfty}(a) compares $C^{Q_x}_{\delta U \delta U} (\omega)$ for porphine, TPP, and TPPL (see SI Figs.~S2(a), S3(a), and S4(a) for $C^{Q_y}_{\delta U \delta U} (\omega)$ of porphine, TPP, and TPPL). This figure shows that, while $C^{Q_x}_{\delta U \delta U} (\omega)$ for porphine and TPP are similar, the spectral density of TPPL shows a greater weight, especially in the region between 600--1400 cm$^{-1}$, which roughly corresponds to internal-ring C-O stretching around 1000 cm$^{-1}$. Because the lactone group contains a carbonyl, one might expect that the vibration of this C=O should modulate the energy gap, leading to a peak near $1800-1900$ $\rm cm^{-1}$. Yet, this peak is conspicuously absent from the TPPL spectral density, while porphine and TPP display greater spectral weight in this region. This is suspect and can indicate a violation of Gaussian statistics. Indeed, employing the Kullback-Leibler divergence between each MD-derived energy gap distribution and their corresponding normal distributions of identical mean and variance (denoted as $D_{\rm KL}^{{\rm MD}|\mathcal{N}}$) to quantify the extent of non-Gaussian behavior, we find that the TPPL $Q_x$ energy gap is more non-Gaussian than both its $Q_y$ counterpart and the energy gaps for porphine and TPP (see SI Sec.~V and Figure~S9). This suggests the need for a closer interrogation of the $Q_x$ energy gap.

Since the only change between TPP and TPPL is the lactone substitution, which largely affects high-frequency intramolecular vibrations, one can interrogate the source of non-Gaussian behavior via a microscopic harmonic analysis of the nuclear motions that modulate the energy gap for an isolated propholactone ring (see Sec.~SVI of the SI). This analysis reveals that the normal modes on the ground electronic surface mix to yield those on the $Q_x$ excited state. Figure~\ref{fig:fig5-Jsd-duscinsky-ginfty}(b) illustrates this by visualizing the Duschinsky rotation, $\mathbf{J}$,\cite{duschinsky1937} of the normal modes upon photoexcitation. The extent to which the Duschinsky matrix displays non-diagonal entries quantifies the amount of mode mixing, which causes non-Gaussian behavior and masks high-frequency vibrational coupling as lower-frequency weight in the spectral density.\cite{Zuehlsdorff2019} Although the Duschinsky matrix is diagonally dominant, indicating that mode mixing is an overall weak effect (see SI Fig.~S11), Fig.~\ref{fig:fig5-Jsd-duscinsky-ginfty}(b) shows a region of the Duschinsky matrix for the $Q_x$ band of porpholactone that reveals non-negligible mode mixing between ground state (horizontal-axis) and excited state (vertical-axis) normal modes with frequencies $> 900$ cm$^{-1}$. 

Does the spurious vibronic progression predicted by our GNCT for the TPPL $Q_x$ band arise from mode mixing-induced non-Gaussianity? To answer this question, we correct the energy gap contribution in our GNCT, $\sigma_{\delta U \delta U} (\omega)$ in Eqs.~\eqref{nature-main-eq:NC-static-spectrum} and \eqref{nature-main-eq:NC-dynamic-spectrum}, with an energy gap contribution that incorporates mode mixing (see SI Sec.~VI). Specifically, we calculate the exact response function for the microscopically harmonic model of the porpholactone ring and add the effect of the collective solvent motions and phenyl torsions in solvated TPPL by extracting the broadening one would obtain from fitting a Debye contribution to the low-frequency portion of our atomistically derived spectral density. Figure~\ref{fig:fig5-Jsd-duscinsky-ginfty}(c) shows the resulting spectrum ($\mathbf{J}^{Q_x}_{\rm PL}$-GNCT) compared to the predicted TPPL $Q$ bands from our GNCT result and the FCHT which yields an unphysical negative absorption feature at $\sim 1.8$ eV. Here, we confirm that by accounting for microscopically harmonic mode mixing corrections in the TPPL $Q_x$ band, our GNCT (purple line) now accurately captures the splitting of both $Q_x$ and $Q_y$ bands of TPPL, resolving all four peaks and agreeing well with the experimental lineshape. Thus, our results for porphine and TPPL confirm our hypothesis that our theory's mechanistic insights enable us to tailor and optimize the optical properties of a molecular system by manipulating nuclear motions that induce statistical transition dipole fluctuations with characteristic timescales.

In summary, we have leveraged our GNCT to successfully simulate, reproduce, and analyze, for the first time, the condensed phase $Q$-band spectra of representative porphyrins: TPP, its parent porphine without phenyl substituents, and the lactone-substituted TPPL. The $Q$ band region plays a key role in the fundamental energy and charge transfer steps in photosynthesis \cite{Porphyrinfundamentals2022, Ma2003, Holt2005, Ruban2007, polívka2010molecular, Bode2009, Son2020}, highlighting the importance of our analysis, which enabled us to settle the long-standing debate over the physical origin of the splitting of the $Q$ bands observed for metal-free porphyrins. We showed that peak-splitting arises from extreme timescale separation in the nuclear motions that modulate selection rules: low-frequency phenyl torsions and high-frequency inner-ring vibrations. The insights we uncovered can be expected to shed light on the dynamical mechanisms at play in metal-substituted porphyrins that often appear in natural and artificial light-harvesting complexes and photo- and electro-catalysis. Such timescale-separated transition dipole fluctuations are also likely crucial in understanding the linewidths and relative intensities between the $Q_x$ and $Q_y$ transitions in the optical spectra of metallo-porphyrins even if the respective bands do not \textit{appear} to split. 

We further demonstrated that our theory can reveal design principles to tune porphyrin optoelectronic behavior. To this end, we exploited the atomistic detail of our GNCT to selectively suppress slow transition dipole fluctuations and tune fast fluctuations of the transition dipoles and energy gap via chemical modifications. By simulating the $Q$ bands of porphine, we proved that severing the pendant phenyls on TPP almost entirely eliminates the low-frequency nuclear motions leading to the systematic suppression of the first and third peaks in the $Q$ band region, in agreement with our theory-led hypothesis. In turn, we also confirmed that substituting the most distant carbons in one of the pyrrole groups in the tetrapyrrole ring indeed shifted the peak height ratios of the second and fourth peaks by tuning the fast transition dipole fluctuations and caused a vibronic progression-induced asymmetry of all peaks, especially those corresponding to the $Q_y$ transition, through the enhancement of energy gap modulations from fast intra-ring motions. 

Our results constitute an exciting breakthrough that opens the door to elucidating the physical mechanisms that produce the versatile optoelectronic properties of the large porphyrin family and beyond. With the promising properties that porphyrins offer as chemically tunable scaffolds to guide energy and charge transfer for energy conversion, catalysis, and even electrochemistry, the ability to employ theory to design chemical structures that exhibit desired functionality is essential. By offering a simple and reliable framework to simulate and interpret the spectral features of chromophores that transcend the standard approximations (e.g., harmonic motions, lack of appreciable non-Condon fluctuations, lack of disorder or thermal effects), our theory sets the stage to provide novel insight, disentangle the interplay between thermally accessible molecular motions and spectral lineshapes, and establish robust design principles for next-generation bioinspired charge and energy transfer technologies in terms of tuning material optoelectronic properties as a function of composite chemical structure. 

\acknowledgements{T.~J.~Z.~acknowledges start-up funding provided by Oregon State University. A.~M.~C.~and Z.~R.~W.~were supported by an Early Career Award in the CPIMS program in the Chemical Sciences, Geosciences, and Biosciences Division of the Office of Basic Energy Sciences of the U.S.~Department of Energy under Award DE-SC0024154. A.~M.~C.~and Z.~R.~W.~thank Josef Michl, David Jonas, Niels Damrauer, Ralph Jimenez and Yuezhi Mao for helpful conversations. Z.~R.~W.~thanks Jo Kurian, Leo Romanetz and Rory McClish for insightful discussions. We collectively thank Tom Sayer and Nanako Shitara for comments on the manuscript.}

\section{Data availability}
The data, input files for MD calculations, excitation energy and transition dipole fluctuations, and input files for generating spectra needed to reproduce our results can be accessed under the following persistent URI: https://doi.org/10.5281/zenodo.13975838.  All molecular spectra systems were calculated using the open-source MolSpeckPy code freely available on GitHub (https://github.com/tjz21/Spectroscopy\_python\_code).

\bibliography{main}

\section{Methods}

\subsection{Theory}
\label{Methods_Sec:Theory-GNCT-nature}

We employ our recently introduced GNCT for predicting the absorption spectra of chromophores in condensed phase systems in the presence of non-Condon effects of arbitrary strength\cite{wiethornGNCTtheory2023}. At the heart of our theory is the recognition that, particularly in the condensed phase, the macroscopic number of nuclear motions that modulate the energy gap $U(\hat{\mathbf{q}})$ and the transition dipole $\boldsymbol{\mu}(\hat{\mathbf{q}})$ cause these quantities to display Gaussian fluctuations around their thermal averages: $U(\hat{\mathbf{q}}) = \langle U \rangle + \delta U(\mathbf{q})$ and $\boldsymbol{\mu}(\hat{\mathbf{q}}) = \langle \boldsymbol{\mu} \rangle + \delta \boldsymbol{\mu}(\hat{\mathbf{q}})$. The reformulation of the problem in terms of Gaussian nuclear variables enables a statistical map to a non-Condon generalized Brownian oscillator model where the ground and excited state PESs can be decomposed into a continuum of harmonic oscillators that shift in equilibrium upon photoexcitation and whose non-Condon fluctuations can be expressed as linear coupling to the oscillator coordinates. This simple rewriting enables us to obtain closed-form expressions for the linear optical response function. 

Our GNCT predicts a spectrum that is completely parameterized by three spectral densities. These encode the nuclear motion-induced fluctuations of the energy gaps and transition dipoles through the temperature-independent Fourier-transformed (imaginary) component of $C(t)$,
\begin{equation}\label{nature-main-eq:antisymmetrized-spectral-density}
    C^{''}_{\delta A \delta B}(\omega) = -i \int_{-\infty}^{\infty} {\rm d} t \, \erm^{i \omega t} {\rm Im} \langle \delta \hat{A}(t) \delta \hat{B} (0) \rangle,
\end{equation}
where ``Im" denotes the imaginary component of the quantum correlation function.
Through the fluctuation-dissipation theorem stating, $C(\omega) = [1 + \coth(\beta \omega/2 )]C^{''}(\omega)$, with $\beta = 1/k_{\rm B} T$, one can recover the full correlation function. The energy gap spectral density, $C^{}_{\delta U \delta U}, (\omega)$, determines the second-order cumulant lineshape function, $g_{\delta U \delta U}(t)$ (see Eq.~8 in Ref.~\onlinecite{wiethornGNCTtheory2023}) that gives rise to $\sigma_{\delta U \delta U} (\omega)$ in Eqs.~\eqref{nature-main-eq:NC-static-spectrum} and \eqref{nature-main-eq:NC-dynamic-spectrum}, 
\begin{equation}
    \sigma_{\delta U \delta U} (\omega) = \int_{-\infty}^{\infty} {\rm d} t \, \erm^{i (\omega - \langle U \rangle)t - g_{\delta U \delta U}(t)},
\end{equation}
where,
\begin{equation}\label{methods_eq:second_cumulant_lineshape_function}
    g_{\delta U \delta U}(t) = \frac{1}{\pi} \int_{0}^\infty {\rm d} \omega \, \frac{C^{''}_{\delta U \delta U} (\omega)}{\omega^2} \Omega(\beta, \omega , t).
\end{equation}
The spectral density $C^{''}_{\delta U \delta U} (\omega)$ encodes all information on the nuclear motion-induced fluctuations of the energy gap specific to the chemical system and $\Omega(\beta, \omega, t)~=~\coth(\beta \omega / 2) [1- \cos(\omega t)] + i [\sin(\omega t) - \omega t]$ is a universal function immutable across chemical problems. Our GNCT, Eq.~\eqref{nature-main-eq:NC-dynamic-spectrum}, also requires two additional spectral densities, $C^{''}_{\delta \boldsymbol{\mu} \delta \boldsymbol{\mu}} (\omega)$ and $\mathbf{C}^{''}_{\delta U \delta \boldsymbol{\mu}} (\omega)$, to parameterize the non-Condon contributions to the total lineshape, 
\begin{subequations}
\begin{align}
    \big[\sigma_{\delta\boldsymbol{\mu} \delta\boldsymbol{\mu}} &\! \ast \! \sigma_{\delta U \delta U} \big](\omega) \nonumber \\=& \int_{-\infty}^{\infty} {\rm d} t \, \erm^{i (\omega - \langle U \rangle)t - g_{\delta U \delta U}(t)} \mathcal{B}_{\delta \boldsymbol{\mu} \delta \boldsymbol{\mu}}(t), \\ \big[\boldsymbol{\sigma}_{\delta U \delta\boldsymbol{\mu}} &\! \ast \! \sigma_{\delta U \delta U} \big] (\omega) \nonumber \\=& \int_{-\infty}^{\infty} {\rm d} t \, \erm^{i( \omega - \langle U \rangle)t - g_{\delta U \delta U}(t)} \boldsymbol{\mathcal{A}}_{\delta U \delta \boldsymbol{\mu}} (t),
\end{align}
\end{subequations}
where
\begin{subequations}
\begin{align}
    \mathcal{B}_{\delta \boldsymbol{\mu} \delta \boldsymbol{\mu}}(t) &= \frac{1}{\pi}\int^{\infty}_{0} {\rm d} \omega \, \frac{C^{''}_{\delta \boldsymbol{\mu} \delta \boldsymbol{\mu}} (\omega)}{\omega^2} \frac{\partial^2}{\partial t^2} \Omega(\beta, \omega, t), \label{methods_eq:second_cumulant_dipole_2nd_moment} \\ \boldsymbol{\mathcal{A}}_{\delta U \delta \boldsymbol{\mu}}(t) &= \frac{i}{\pi}\int^{\infty}_{0} {\rm d} \omega \, \frac{\mathbf{C}^{''}_{\delta U \delta \boldsymbol{\mu}} (\omega)}{\omega^2} \frac{\partial}{\partial t} \Omega(\beta, \omega, t)\label{methods_eq:second_cumulant_energy_dipole_joint_moment}.
\end{align}
\end{subequations}
and $C^{''}_{\delta \boldsymbol{\mu} \delta \boldsymbol{\mu}} (\omega)$ and $\mathbf{C}^{''}_{\delta U \delta \boldsymbol{\mu}} (\omega)$ encode all information on how nuclear motion-induced fluctuations of the transition dipoles impact spectra. 

The spectral densities in Eqs.~\eqref{methods_eq:second_cumulant_lineshape_function}, \eqref{methods_eq:second_cumulant_dipole_2nd_moment} and \eqref{methods_eq:second_cumulant_energy_dipole_joint_moment} arise from the Fourier transform of quantum equilibrium time correlation functions (see Eqs.~(10), (19) and (20) in Ref.~\onlinecite{wiethornGNCTtheory2023}). Since exact quantum correlation functions are generally inaccessible in condensed phase systems, we employ the Kubo (harmonic) quantum correction factor to approximate quantum correlation functions with their classical counterparts \cite{Bader1994, Egorov1999, Kim2002b, mano2004quantum, ramirez2004quantum},
\begin{subequations}\label{eq:spectral-densities-in-terms-of-classical-MD}
\begin{align}
    C^{''}_{\delta U \delta U} (\omega) &\approx \theta(\omega) \frac{\beta \omega}{2} \int_{-\infty}^{\infty} \text{d}t \, \erm^{i \omega t} \langle \delta U( t) \delta U( 0) \rangle_{\rm cl}, \\ C^{''}_{\delta \boldsymbol{\mu} \delta \boldsymbol{\mu}} (\omega) &\approx \theta(\omega) \frac{\beta \omega}{2} \int_{-\infty}^{\infty} \text{d}t \, \erm^{i \omega t} \langle \delta \boldsymbol{\mu}( t) \mdot \delta \boldsymbol{\mu}( 0) \rangle_{\rm cl}, \\
    \mathbf{C}^{''}_{\delta U \delta \boldsymbol{\mu}} (\omega) &\approx  \theta(\omega) \frac{\beta \omega}{2} \int_{-\infty}^{\infty} \text{d}t \, \erm^{i \omega t} \langle \delta U( t) \delta \boldsymbol{\mu}(0) \rangle_{\rm cl}.
\end{align}
\end{subequations}

\subsection{Obtaining the spectral densities from atomistic simulation}
\label{Methods_Sec:Spectral-densities-from-MD-GNCT-nature}
We construct the equilibrium time correlation functions from mixed quantum mechanics/molecular mechanics (QM/MM) MD simulations of the chromophores in CS$_2$. In the following subsections, we detail the necessary steps to calculate spectra from MD simulations paired with electronic structure calculations.

\subsubsection{Mixed quantum-classical simulations}
\label{Methods_Sec:MD-simulations-GNCT-nature}

We use AmberTools\cite{amber} to simulate the dynamics of porphine, TPP, and TPPL using mixed quantum mechanics/molecular mechanics (QM/MM) MD simulations in 30~\AA\ spheres of CS$_2$ solvent. To reduce the computational costs associated with long equilibration times in such calculations, a periodic box of 989 CS$_2$ molecules was first heated to 300~K and equilibrated at 1~atm over a total of 520~ps of MM-only dynamics, with a time step of 2~fs. The chromophores were then equilibrated in the center of CS$_2$ spheres in open boundary conditions for 50~ps of MM-only NVT dynamics using a Langevin thermostat with a collision frequency of 1~ps$^{-1}$ and a time step of 0.5~fs. Following the MM simulations while holding the same parameters (timestep, thermostat collision frequency, box size and dimension), we performed 22~ps (for TPPL) and 42~ps (for porphine and TPP) of equilibrium simulations while treating the chromophores quantum mechanically. To perform QM/MM simulations, we employ the interface between Amber20 and TeraChem\cite{Isborn2012} software packages for molecular mechanical and quantum mechanical calculations, respectively. The electronic structure and forces of the embedded quantum mechanical chromophore were calculated with density functional theory (DFT) at the level of CAM-B3LYP\cite{Yanai2004} functional with the 6-31+G*\cite{Dunning1990} basis set. The forces of the solvent environment (and the chromophore in equilibration trajectories) were calculated using the generalized Amber force field (GAFF) as parameterized by the AmberTools utilities. The choice of a full QM description of the chromophore during the MD was made to avoid known issues arising from mismatches in the Hamiltonians used for propagation and calculations of vertical excitation energies when constructing equilibrium quantum correlation functions\cite{Zwier2007, Rosnik2015, Chandraskaran2015, Lee2016, Andreussi2017}. The first 2~ps of QM/MM dynamics were discarded to allow for additional equilibration after switching from the MM to the QM Hamiltonian for the chromophore, yielding 40~ps (for TPP and porphine) and 20~ps (for TPPL) of usable trajectory. 

\subsubsection{Building classical correlation functions from quantum mechanical calculations}

We construct the classical correlation functions in Eq.~\ref{eq:spectral-densities-in-terms-of-classical-MD} using the open-source package MolSpeckPy\cite{MolSpecPycode} by sampling the equilibrium QM/MM trajectories every 2~fs (for a total of 40 ps for TPP and porphine and 20 ps for TPPL) to calculate the vertical excitations and transition dipoles between the first two excited states of each porphyrin ($Q_x$ and $Q_y$). We choose a sampling interval of 2~fs to accurately capture the influence of both slow and fast molecular vibrations on the energy gaps and transition dipoles, and employ time-dependent density-functional theory (TDDFT) to compute both the vertical excitation energies and transition dipoles along the MD trajectories of TPP, porphine and TPPL. All calculations were performed using full TDDFT, without resorting to the simplified Tamm-Dancoff approximation\cite{Hirata1999} using the same level of theory and basis set as those used to run the QM/MM simulations. 

\subsubsection{Rotating transition dipole correlation functions into a common reference frame}

In MD simulations, solvated chromophores are often free to rotate. These motions can contaminate classical correlation functions of vector quantities, such as the transition dipole, which can lead to erroneous vector norms if they are not calculated with respect to a consistent center of mass and reference frame. To remove the effect of rotations, we take the optimized geometry of the chromophore in vacuum as a reference geometry and rotate and translate the chromophore coordinates in each MD snapshot to minimize the mass-weighted square displacement (MWSD) of the MD coordinates from the reference geometry coordinates. Minimizing the MWSD is equivalent to the so-called Eckart conditions\cite{Eckart1935}, and yields chromophore coordinates along the entire MD trajectory that are free from rotational and translational contributions. 

In defining the rotation matrix to obtain the chromophore coordinates in the new reference frame, we follow the algorithm outlined in Ref.~\onlinecite{Krasnoshchekov2014}. First, both the reference geometry and the geometry of a given MD snapshot are transformed into the center-of-mass frame with respect to the chromophore coordinates only. The 3D rotation matrix ${\textrm{\textbf{U}}}$ rotating the MD snapshot coordinates into the reference frame is then given by, 
\begin{widetext}
\begin{equation}
    \textrm{\textbf{U}}=
    \begin{pmatrix}
    \left(q_0^2+q^2_1-q^2_2-q^2_3\right) & 2\left(q_1q_2+q_0q_3\right)& 2\left(q_1q_3-q_0q_2\right)\\
    2\left(q_1q_2-q_0q_3\right)& \left(q_0^2-q^2_1+q^2_2-q^2_3\right)& 2\left(q_2q_3+q_0q_1\right)\\
    2\left(q_1q_3+q_0q_2\right)&2\left(q_2q_3-q_0q_1\right) &\left(q_0^2-q^2_1-q^2_2+q^2_3\right)
    \end{pmatrix}.
\end{equation}
Here, $\textbf{\textrm{q}}=\left[q_0,q_1,q_2,q_3\right]$ is the eigenvector with the lowest eigenvalue of the 4x4 symmetric matrix $\textbf{\textrm{C}}$
\begin{equation}
    \textrm{\textbf{C}}=
    \sum_a^N m_a \begin{pmatrix}
    \left(x_{-a}^2+y_{-a}^2+z_{-a}^2 \right) & \left(y_{+a}z_{-a}-y_{-a}z_{+a} \right) & \left(x_{-a}z_{+a}-x_{+a}z_{-a} \right) & \left(x_{+a}y_{-a}-x_{-a}y_{+a} \right) \\
    
    \left(y_{+a}z_{-a}-y_{-a}z_{+a} \right) & \left(x_{-a}^2+y_{+a}^2+z_{+a}^2 \right) & \left(x_{-a}y_{-a}-x_{+a}y_{+a} \right) & \left(x_{-a}z_{-a}-x_{+a}z_{+a} \right) \\
    
    \left(x_{-a}z_{+a}-x_{+a}z_{-a} \right) & \left(x_{-a}y_{-a}-x_{+a}y_{+a} \right)& \left(x_{+a}^2+y_{-a}^2+z_{+a}^2 \right) & \left(y_{-a}z_{-a}-y_{+a}z_{+a}\right)\\
    
    \left(x_{+a}y_{-a}-x_{-a}y_{+a} \right) & \left(x_{-a}z_{-a}-x_{+a}z_{+a} \right) & \left(y_{-a}z_{-a}-y_{+a}z_{+a}\right) & \left(x_{+a}^2+y_{+a}^2+z_{-a}^2 \right)
    \end{pmatrix}.
\end{equation}
\end{widetext}
$N$ denotes the number of atoms and $m_a$ is the mass of atom $a$. We have introduced the notation of $x_{+a}=x_{a,\textrm{ref}}+x_{a}$ and $x_{-a}=x_{a,\textrm{ref}}-x_{a}$, where $x_{a,\textrm{ref}}$ denotes the $x$-coordinate of the $a$-th atom of the reference geometry in the center of mass frame, and $x_{a}$ denotes the $x$-coordinate of the $a$-th atom of the given MD geometry that is to be transformed into the Eckart frame, again in the center of mass frame. Constructing the rotation matrix for a given MD snapshot thus reduces to calculating the matrix elements of $\textbf{\textrm{C}}$ and computing its lowest eigenvalue and corresponding eigenvector. 

Transforming all transition dipole moments along the MD trajectory into a consistent Eckart frame before constructing classical correlation functions then guarantees that the correlation functions obtained only contain contributions due to vibrational motion of the chromophore and the environment, not rotational  motion. We have found that the approach yields correlation functions whose Fourier representations show well-defined peaks at frequencies corresponding to identifiable vibrational modes of the chromophore (see SI Sec.~III), while still retaining the influence of interactions with the complex solvent environment. For all calculations presented in this work, the Eckart frame is chosen such that the prophyrin ring is in the $x-y$ plane, with the central N-H bonds aligned along the $x$-axis. 

While the Eckart approach fixes the direction of the transition dipole to a common reference frame, it does not treat a remaining ambiguity in the transition dipole phase. Specifically, in contrast to the transition dipole magnitude, $|\hat{\boldsymbol{\mu}}|^2$, the transition dipole vector is itself not a physical observable. This means that the transition dipole is defined to within an arbitrary phase of $\pm 1$. Hence, each transition dipole component ($\hat{\mu}_x$, $\hat{\mu}_y$, or $\hat{\mu}_z$) can arbitrarily take its positive or negative values and, as a result, the sign of the transition dipole can flip sign between subsequent simulation frames. This arbitrary sign-flipping leads to raw transition dipole trajectories obtained through a direct MD sampling of the ground PES that contain jump discontinuities that contaminate the correlation functions, and therefore, spectral densities with high-frequency regions displaying over-saturated spectral weight. To address this issue in the present work, we perform the following steps:
\begin{itemize}
    \item At the ground state optimized geometry of the chromophore in vacuum, $\textbf{R}_\textrm{GS}\equiv\{\textbf{r}^\textrm{GS}_i\}$, we compute the derivative of the transition dipole moment $\left. \frac{\partial \boldsymbol{\mu}^{Q_{x/y}}}{\partial \textbf{r}_i}\right|_{\textbf{R}_\textrm{GS}}$ with respect to nuclear coordinates $\{\textbf{r}_i\}$ for both the $Q_x$ and the $Q_y$ transition using a finite difference scheme. 
    \item 
    For a given MD snapshot with chromophore coordinates $\textbf{R}_j$ in the Eckart frame, we compute $\Delta \textbf{R}_j=\textbf{R}_j-\textbf{R}_0$, the deviation of the nuclear coordinates from their equilibrium position. 
    \item 
    We compute $\delta \boldsymbol{\mu}^\textrm{HT}_{j}=\left. \frac{\partial \boldsymbol{\mu}^{Q_{x/y}}}{\partial \textbf{r}_i}\right|_{\textbf{R}_\textrm{GS}}\, \Delta \textbf{R}_j$, corresponding to the dipole fluctuations predicted for a given MD snapshot $j$ based on a Taylor expansion to first order at the ground state equilibrium geometry.
    \item 
    The sign of the transition dipole moment fluctuations computed directly with TDDFT for the given MD snapshot is chosen to maximize the overlap with $\delta \boldsymbol{\mu}^\textrm{HT}_{j}$. 
\end{itemize}

\label{Methods_Sec:Normal-modes-GNCT-nature}

For all systems studied in this work, we compute ground state vibrational modes of the respective molecule using density-functional theory (DFT) as implemented in the Gaussian package \cite{g16}, with the same calculation parameters as used in the QM/MM dynamics (see SI Sec.~II~C). Normal mode frequencies are not expected to align exactly with peaks in the spectral densities, as the MD sampling of the full potential energy surfaces accounts for anharmonic effects. However, for the systems studied in this work, transition dipole and energy gap spectral densities $C_{\delta\boldsymbol{\mu} \delta\boldsymbol{\mu}}(\omega)$ and $C_{\delta U  \delta U}(\omega)$ generally show sharp, well-defined, peaks, in most cases allowing for an unambiguous assignment to specific vibrational normal modes. The lowest frequency peaks of $C_{\delta\boldsymbol{\mu} \delta\boldsymbol{\mu}}(\omega)$ cannot be easily assigned to a single vibrational mode, but we instead observe that all the low-frequency modes of TPP and TPPL have some similar characteristics. A detailed normal mode analysis of all systems studied is provided in SI Sec.~III.

\subsection{Model system calculations}
\label{Methods_Sec:Model-system-spectral-densities-GNCT-nature}

To elucidate the microscopic mechanism responsible for the $Q$-band splitting in porphyrin spectra, we systematically tested which basic features of the MD-based spectral densities, and particularly $C_{\delta\boldsymbol{\mu} \delta\boldsymbol{\mu}}(\omega)$, are required to recapitulate the observed spectra. This enabled us to produce minimal but physically transparent spectral densities that elucidate that the timescale separation in $C_{\delta\boldsymbol{\mu} \delta\boldsymbol{\mu}}(\omega)$ causes the splitting of the $Q$-bands. To arrive at physically transparent spectral densities from the MD-derived ones, we employed a Debye spectral density of the form, 
\begin{equation}
    C^{''}_{\delta \boldsymbol{\mu} \delta \boldsymbol{\mu}} (\omega_{\rm lf}) = 2 d \omega_c \frac{\omega}{\omega^2 + \omega_c},
\end{equation}
to account for collective solvent motions and anharmonic intramolecular motions (i.e., the TPP phenyl torsions). Here, the quantities $d$ and $\omega_c$ correspond to the maximum strength (amplitude) of the average frequency (center frequency) of the low-frequency transition-dipole fluctuations. In our analysis, we set $d^{Q_x} \sim 3.68 \times 10^{-4} \dwn$ and $d^{Q_y} \sim 5.30 \times 10^{-4} \dwn$, and $\omega^{Q_x}_c \sim 109.78 \wn$ and $\omega^{Q_y}_c \sim 76.82 \wn$ to match the values of the single peak in the low-frequency region of the MD-based $K_{\rm S_1} (\omega)$. To capture the high-frequency structure of $C_{\delta\boldsymbol{\mu} \delta\boldsymbol{\mu}}(\omega)$ without loss of qualitative agreement in the resulting TPP ${\rm S}_0 \to {\rm S_1}$ $Q$-band spectrum, we subsumed much of the complexity into a single Gaussian envelope,
\begin{equation}
     C^{''}_{\delta \boldsymbol{\mu} \delta \boldsymbol{\mu}} (\omega_{\rm hf}) = \frac{D}{\sigma \sqrt{2 \pi}} \erm^{-\frac{(\omega - \Omega_c)^2}{2 \sigma^2}}.
\end{equation}
Here, the quantities $D$ and $\Omega_c$ correspond to the maximum strength (amplitude) of the average frequency (center frequency) of the high-frequency transition-dipole fluctuations, where $\sigma$ is their standard deviation.

To construct a Gaussian profile that faithfully represents the inner-pyrrole ring vibrations of TPP, we extracted the mean and standard deviations from a range of frequencies that capture the high-frequency region of the MD-based $C^{''}_{\delta \boldsymbol{\mu} \delta \boldsymbol{\mu}}(\omega)$. We determined the range of $1159.91-2868.70 \wn$ to sufficiently sample this high-frequency region providing $\Omega^{Q_x}_c \sim 1522.37 \wn$ and $\Omega^{Q_y}_c \sim 1542.94 \wn$ as the respective center frequencies, and $\sigma^{Q_x} \sim 204.73 \wn$ and $\sigma^{Q_y} \sim 124.07 \wn$ as the respective variances. Lastly, we chose $D^{Q_x} \sim 1.86 \times 10^{-2} \dwn$ and $D^{Q_y} \sim 6.19 \times 10^{-2} \dwn$ such that the area under our Gaussian profiles matched the area under the chosen high-frequency region of the MD-Based spectral densities.

\subsection{Non-Gaussian energy gap fluctuations in TPPL: corrections from harmonic analysis}

We test the influence of non-Gaussian energy gap fluctuations induced by vibrational mode mixing in the $Q_x$ band of TPPL by employing a microscopically harmonic analysis for porpholactone (PL), i.e., TPPL without the phenyl substituents in the gas phase. The insight that makes this possible is the separation of timescales of the nuclear motions that modulate the energy gap and transition dipoles, suggesting that one may approximately interrogate he influence of the low- and high-frequency motions on the energy gap \textit{separately} via distinct theoretical treatments. Specifically, we make the following substitution to Eqs.~\eqref{nature-main-eq:NC-static-spectrum} and \eqref{nature-main-eq:NC-dynamic-spectrum},
\begin{equation}
    \sigma_{\delta U \delta U } (\omega) \to \sigma^{\rm mix}_{\infty} (\omega) = \int_{-\infty}^{\infty} {\rm d}t \, \erm^{i (\omega - \langle U \rangle)t - G^{\rm mix}_{\infty}(t)},
\end{equation}
where, $G^{\rm mix}_{\infty}(t)$ is a modified lineshape function with mixed low- and high-frequency contributions,
\begin{equation}
    G^{\rm mix}_{\infty}(t) = g_{\delta U \delta U}^{\rm TPPL}(t; \omega_{lf}) + g_{\delta U^{\infty}}^{\rm FC-PL}(t; \omega_{\rm hf}).
\end{equation}
We use $\omega_{\rm lf}$ and $\omega_{\rm hf}$ to denote that these lineshape functions are parameterized by low- or high-frequency nuclear motions, respectively. $g_{\infty}^{\rm FCPL}(t; \omega_{\rm hf})$ is the infinite-order cumulant lineshape function for porpholactone obtained via a Franck-Condon harmonic analysis (FCPL). We calculate ${\rm e}^{- g_{\infty}^{\rm FCPL} (t)}$ by solving directly for the optical response function in the generalized Brownian oscillator model (GBOM) by invoking a normal-mode representation for the ground-state vibrational wavefunctions and employing the Duschinsky rotation matrix to account for non-Gaussian mode mixing between ground and excited state normal modes (See SI Sec.~VI), as outlined in Ref.~\onlinecite{Zuehlsdorff2019}. 

Due to the harmonic approximation, $g_{\infty}^{\rm FCPL} (t)$ is only stable for high-frequency vibrations. To incorporate the low-frequency collective solvent motions native to the condensed phase and the anharmonic torsions of the TPPL pendant phenyls, we fit a Debye spectral density to the low-frequency region of $C_{\delta U_x \delta U_x}$ predicted by our GNCT (See SI Sec.~VI). Using this model spectral density, we calculate $g_2^{\rm TPPL}(t; \omega_{lf})$, which amounts to neglecting the high-frequency detail in $C_{\delta U_x \delta U_x}$ of TPPL. 

\subsection{Spectra generation}

The GNCT approach described in this work is implemented in MolSpeckPy, an open-source Python package freely available on GitHub\cite{MolSpecPycode}. All molecular spectra shown in this work are calculated using the Python package with correlation functions computed up to  500~fs in the time domain. To guarantee well-behaved Fourier transforms in Eq.~(\ref{eq:spectral-densities-in-terms-of-classical-MD}), all classical correlation functions are multiplied with a decaying exponential with $t_{1/2}=500$~fs. A timestep of $\Delta t=0.167$~fs is used throughout. All input files and raw data necessary to reproduce the data presented in this work are accessible via the following persistent URI: https://doi.org/10.5281/zenodo.13975838.

Full absorption lineshapes are created by adding linear optical spectra computed for the $Q_x$ and the $Q_y$ transition, following the assumption that the influence of nonadiabatic couplings between the transitions is negligible (see SI Sec.~IV). The CAM-B3LYP functional consistently underestimates the energy separation of $Q_x$ and $Q_y$ transitions in the Condon region for all systems studied in this work. To compare directly with experiment, linear spectra due to the $Q_x$ transition of porphine, TPP and TPPL are shifted by -0.15 eV, -0.155 eV, and -0.17 eV, respectively whereas the $Q_y$ transition of porphine, TPP and TPPL are shifted by -0.028 eV, -0.018 eV, and 0 eV, respectively, before adding the lineshapes.

\end{document}